\documentclass[a4paper]{PoS}

\usepackage{bm}
\usepackage{amsmath}
\usepackage{mathtools}
\usepackage{siunitx}
\usepackage{braket}

\usepackage{booktabs}
\usepackage[abbreviations,british]{foreign}

\title{The \bm{$H$} dibaryon from lattice QCD with SU(3) flavor symmetry}

\ShortTitle{The $H$ dibaryon from lattice QCD with SU(3) flavor symmetry}

\author{\speaker{Andrew~Hanlon}~$^a$, Anthony~Francis$^b$, Jeremy~Green$^c$, Parikshit~Junnarkar$^d$, Hartmut~Wittig$^{ae}$\\
\llap{$^a$}Helmholtz-Institut Mainz, Johannes Gutenberg-Universit\"at, 55099 Mainz, Germany\\
\llap{$^b$}Theoretical Physics Department, CERN, CH-1211 Geneva 23, Switzerland\\
\llap{$^c$}NIC, Deutsches Elektronen-Synchrotron, D-15738 Zeuthen, Germany\\
\llap{$^d$}Tata Institute of Fundamental Research (TIFR), 1 Homi Bhabha Road, Mumbai 400005. India\\
\llap{$^e$}PRISMA Cluster of Excellence and Institut f\"ur Kernphysik, Johannes Gutenberg-Universit\"at, 55099 Mainz, Germany\\
E-mail: \email{ahanlon@uni-mainz.de}, \email{anthony.francis@cern.ch}, \email{jeremy.green@desy.de}, \email{parikshit@theory.tifr.res.in}, \email{hartmut.wittig@uni-mainz.de}}

\abstract{We show recent results from the Mainz group using $N_f = 2 + 1$ CLS ensembles generated at the $SU(3)$ symmetric point.
  Temporal correlation functions using two-baryon interpolating operators are calculated with the distillation method.
  In addition to the spin-0 operators relevant for studying the $H$ dibaryon, we added spin-1 operators to our basis,
  thereby extending our results to other flavor sectors.
  These preliminary results show a finite-volume energy below the $\Lambda \Lambda$ threshold.
  Further calculations are necessary to establish whether the $H$ dibaryon is bound at the physical point.} 
  
\FullConference{The 36th Annual International Symposium on Lattice Field Theory - LATTICE2018\\
		22-28 July, 2018\\
		Michigan State University, East Lansing, Michigan, USA.}

\begin{document}

\section{Introduction}

Searches for the elusive $H$ dibaryon,\footnote{
  For a recent review on the status of dibaryon searches, including the $H$ dibaryon, see Ref.~\cite{dibaryon_review}}
an $SU(3)$ flavor singlet with $J^P = 0^+$ and quark content $uuddss$,
have persisted since its prediction by Jaffe in 1977~\cite{jaffe_prediction} as a deeply bound state of a two $\Lambda$ baryon system.
Various theoretical model calculations predict a wide spread of masses, ranging from deeply bound to unbound with respect to the $\Lambda \Lambda$ threshold~\cite{dibaryon_review}.
Constraints from the ``NAGARA'' event have given an upper limit for the $H$-dibaryon binding energy of approximately \SI{7}{MeV} at the $90\%$ confidence level~\cite{experiment},
which is drastically smaller than the original bag model prediction of around \SI{80}{MeV}.

Lattice results with $N_f = 3$ dynamical quarks have been performed by the NPLQCD~\cite{NPLQCD_Nf3} and HAL QCD~\cite{HALQCD_Nf3} collaborations,
which both indicate the existence of a bound $H$ dibaryon but with drastically different binding energies for $m_\pi \approx \SI{800}{MeV}$.
Further results with $N_f = 2 + 1$ dynamical quarks have also been performed by these two collaborations,
with NPLQCD reporting evidence for a bound $H$ dibaryon at pion masses of \SI{230}{MeV} and \SI{390}{MeV}~\cite{NPLQCD_Nf21},
and HAL QCD indicating that the $H$ dibaryon may be a $\Lambda \Lambda$ resonance using nearly physical quark masses~\cite{HALQCD_Nf21}.

Recent lattice results from the Mainz group using $N_f = 2$ ensembles with a quenched strange quark 
also demonstrate a bound $H$ dibaryon for an $SU(3)$-symmetric and $SU(3)$-broken setup with pion masses of \SI{960}{MeV} and \SI{440}{MeV}, respectively~\cite{mainz_2flavor}.
These results also include an analysis based on L\"uscher's finite-volume quantization condition~\cite{luescher,luescher_moving} in the $SU(3)$-symmetric case to assess the finite-volume effects there.
In the following, we describe the recent progress from the Mainz group on extensions to $N_f = 2 + 1$ ensembles at the $SU(3)$ symmetric point.

\section{Lattice Methodology}

The gauge ensembles used in this work, shown in Table~\ref{tab:ensembles}, were generated as a part of the Coordinated Lattice Simulations (CLS) $N_f = 2 + 1$ effort~\cite{ensembles}.
These ensembles use non-perturbatively $O(a)$-improved Wilson fermions and the tree-level $O(a^2)$ improved L\"{u}scher-Weisz gauge action.
Although many of the ensembles are generated with open boundary conditions in time in order to avoid the freezing of the global topological charge on fine lattices,
we also make use of an ensemble with periodic boundary conditions in time for the gauge fields.
The rationale for our choice of ensembles was from an interest in the finite-volume dependence, as each of these ensembles have different volumes but are otherwise similar.

As was demonstrated in Ref.~\cite{mainz_2flavor},
the distillation method~\cite{distillation} offers a substantial improvement to the quality of data at a similar cost compared to the use of smeared point sources.\footnote{
  The cost estimate is based on the number of inversions required between the two methods.
  For a more accurate comparison, one should take the costs of contractions into consideration, which are generally larger for distillation scaling as $N^4_{\rm LapH}$.}
Therefore, we exclusively use distillation on the $N_f = 2 + 1$ ensembles.
This method takes advantage of Laplacian Heaviside (LapH) smearing applied to the quark fields,
allowing for a projection of the quark propagator onto the $N_{\rm LapH}$ lowest modes of the gauge-covariant Laplacian
and thereby reducing the number of inversions required to compute timeslice-to-all propagators.
In order to determine the number of eigenvectors of the Laplacian needed, a comparison of the effective energy for an octet baryon was made with various values of $N_{\rm LapH}$.
As $N_{\rm LapH}$ becomes small, the statistical error increases, but the plateau also occurs at an earlier timeslice. 
Thus a comparison is made between the statistical error on the effective energies relative to their respective plateau onset.
This allows for a relatively modest number of eigenvectors.

\begin{table}[t]
  \begin{center}
    \begin{tabular}{ccccccccc}
      \toprule
      id &   $\beta$ &  $N_\mathrm{s}$  &  $N_\mathrm{t}$  &  $m_\pi$[MeV] & $N_{\rm conf}$ & $N_{\rm LapH}$ & $N_{\rm tsrc}$ & BC \\
      \midrule
      U103 & 3.40 & 24 & 128 & 420 & 5721 & 20 & 5 & open     \\
      H101 & 3.40 & 32 & 96  & 420 & 2016 & 48 & 4 & open     \\
      B450 & 3.46 & 32 & 64  & 420 & 1612 & 32 & 4 & periodic \\
      \bottomrule
    \end{tabular}
    \caption{\label{tab:ensembles}The $N_f = 2 + 1$ CLS ensembles used in this work.
      $N_{\rm LapH}$ is the number of modes of the gauge-covariant Laplacian used in the method of distillation,
      and $N_{\rm tsrc}$ is the number of source timeslices used.
      The boundary conditions refer to those used for the gauge fields in the temporal direction.}
  \end{center} 
\end{table}

\section{Interpolating Operators}

The original bag model prediction by Jaffe described a tightly bound six-quark color singlet hadron,
which qualitatively resemble the hexaquark operators used in Ref.~\cite{mainz_2flavor}.
However, it was found in that study that these hexaquark operators had slower ground-state saturation as compared to two-baryon operators,
and that by the time the plateau had been reached the noise did not allow for a statistically significant shift from the $\Lambda \Lambda$ threshold to be determined.
One could qualitatively explain this poor overlap of the hexaquark operators onto the ground state as being due to a small binding energy,
and thus the $H$ dibaryon may more closely resemble a loosely bound two-baryon state.

For the reasons stated above, and due to the high contraction cost for hexaquark operators in the distillation method,
we only consider baryon-baryon operators in this work.
These operators are constructed from individually momentum-projected octet baryon operators of the form
\begin{equation}
  B_\alpha (\bm{p},t) [uvw] \equiv \sum_{\bm{x}} e^{-i \bm{p} \cdot \bm{x}} \epsilon_{abc} (v^a C \gamma_5 P_+ w^b) u^c_\alpha ,
  \label{eq:octed_baryons}
\end{equation}
where $C$ is the charge conjugation operator and $P_+ = \frac{1+\gamma_0}{2}$ is a projector to positive parity.
From these single-baryon operators, spin-zero and spin-one combinations can be constructed respectively in the following way
\begin{subequations}
\begin{align}
  [B_1 B_2]_0 (\bm{p}_1, \bm{p}_2) &= B^{(1)} (\bm{p}_1) C \gamma_5 P_+ B^{(2)} (\bm{p}_2) , \\
  [B_1 B_2]_i (\bm{p}_1, \bm{p}_2) &= B^{(1)} (\bm{p}_1) C \gamma_i P_+ B^{(2)} (\bm{p}_2) .
\end{align}
\label{eq:two_baryon_operators}
\end{subequations}
For the $H$-dibaryon sector, we need to form $I=0$, $S=-2$ spin-0 operators that are flavor-symmetric (the anti-symmetric flavor combinations do not contribute to $J^P = 0^+$).
This can be done with $\Lambda \Lambda$, $\Sigma \Sigma$, and $N \Xi$ operators,\footnote{
  For the exact expressions, see Ref.~\cite{mainz_2flavor}.}
which can further be combined so as to transform according to definite irreducible representations (irreps) of $SU(3)$ flavor~\cite{su3}.
Although the $H$ dibaryon lives in the $\bm{1}$-dimensional irrep, this irrep will mix with the $\bm{27}$- and $\bm{8}$-dimensional irreps upon $SU(3)$ symmetry breaking,
and therefore it is important to study each of them, even in the $SU(3)$ symmetric case.
We have also computed correlators involving the spin-1, $I=1$, and flavor anti-symmetric operators, with preliminary results shown in the top right of Fig.~\ref{fig:ground_states}.

Lastly, due to the reduced rotational symmetry of a cube, it is important to make sure our operators transform irreducibly under the lattice symmetry group.
To this end, a \textsc{Python} package was developed to ensure the correct transformation properties of all our operators.

\section{Results}

\begin{figure}[t]
  \begin{center}
    \includegraphics[width=0.5\textwidth]{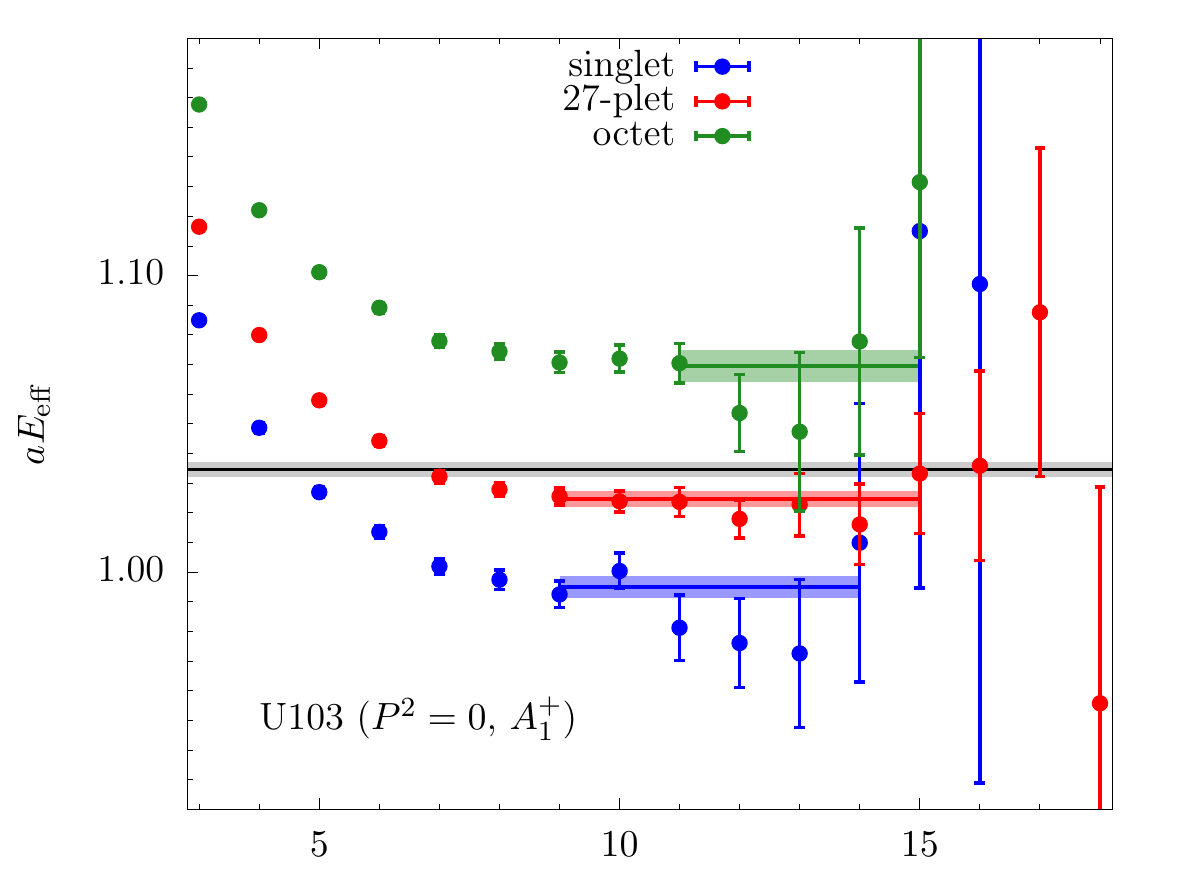}
    \hspace{-0.6cm} \includegraphics[width=0.5\textwidth]{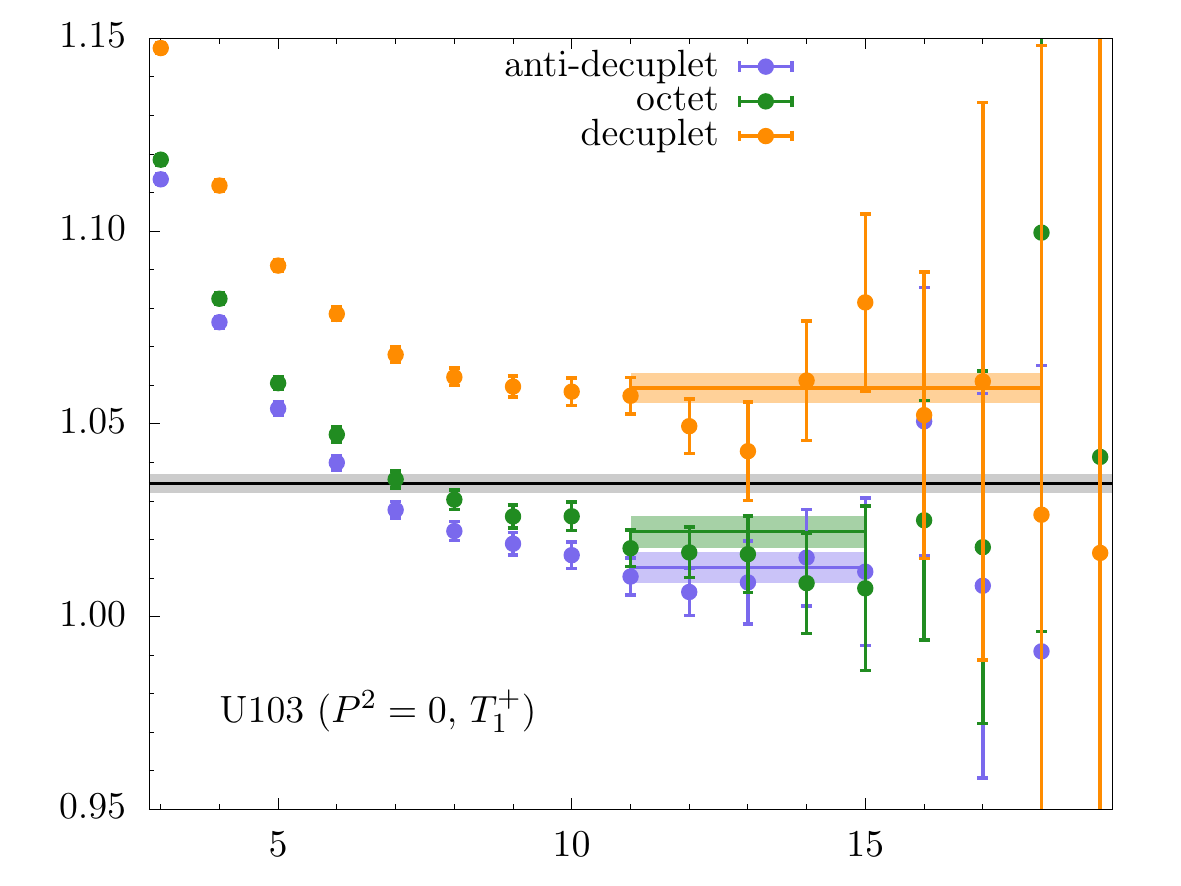}\\[-0.2cm]
    \includegraphics[width=0.5\textwidth]{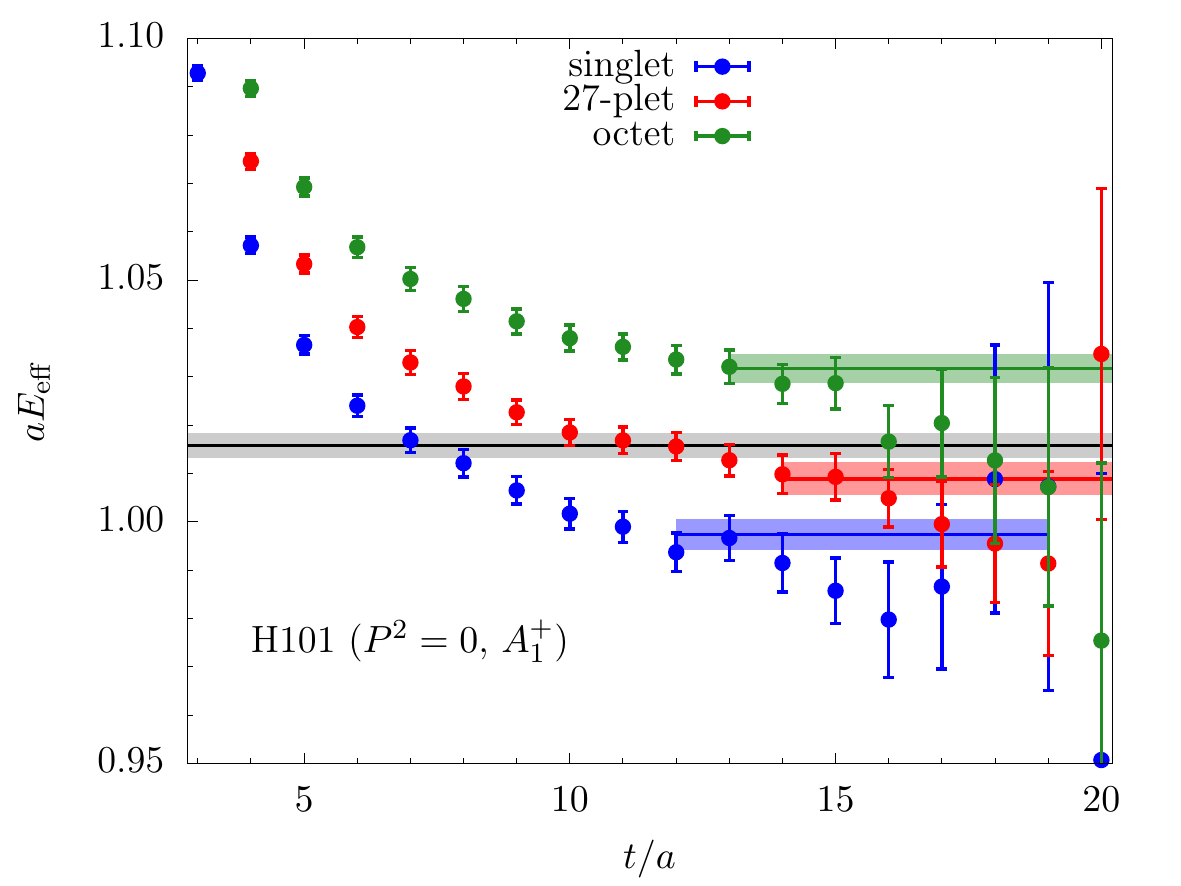}
    \hspace{-0.6cm} \includegraphics[width=0.5\textwidth]{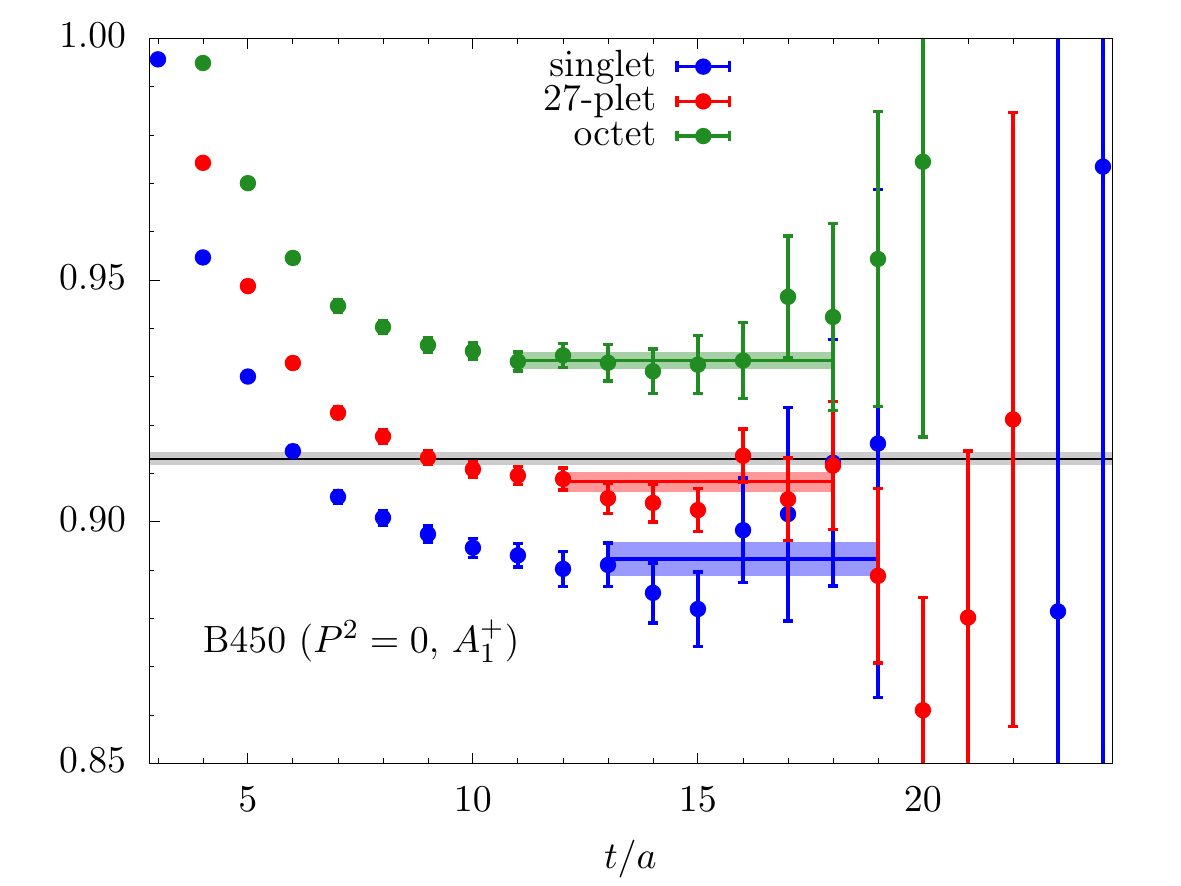}
    \caption{\label{fig:ground_states}The ground state energies at rest in different irreps of $SU(3)$ flavor.
      Each of these effective energies corresponds to the lowest energy extracted from a diagonalized $3 \times 3$ correlation matrix.
      The top row uses ensemble U103 with spin-0 operators on the left and spin-1 operators on the right.
      The bottom row uses only spin-0 operators with ensemble H101 on the left and B450 on the right.
      The black line with grey error band indicates the two-octet-baryon threshold.}
  \end{center}
\end{figure}

\begin{figure}[t]
  \begin{center}
                     \includegraphics[width=0.3925\textwidth]{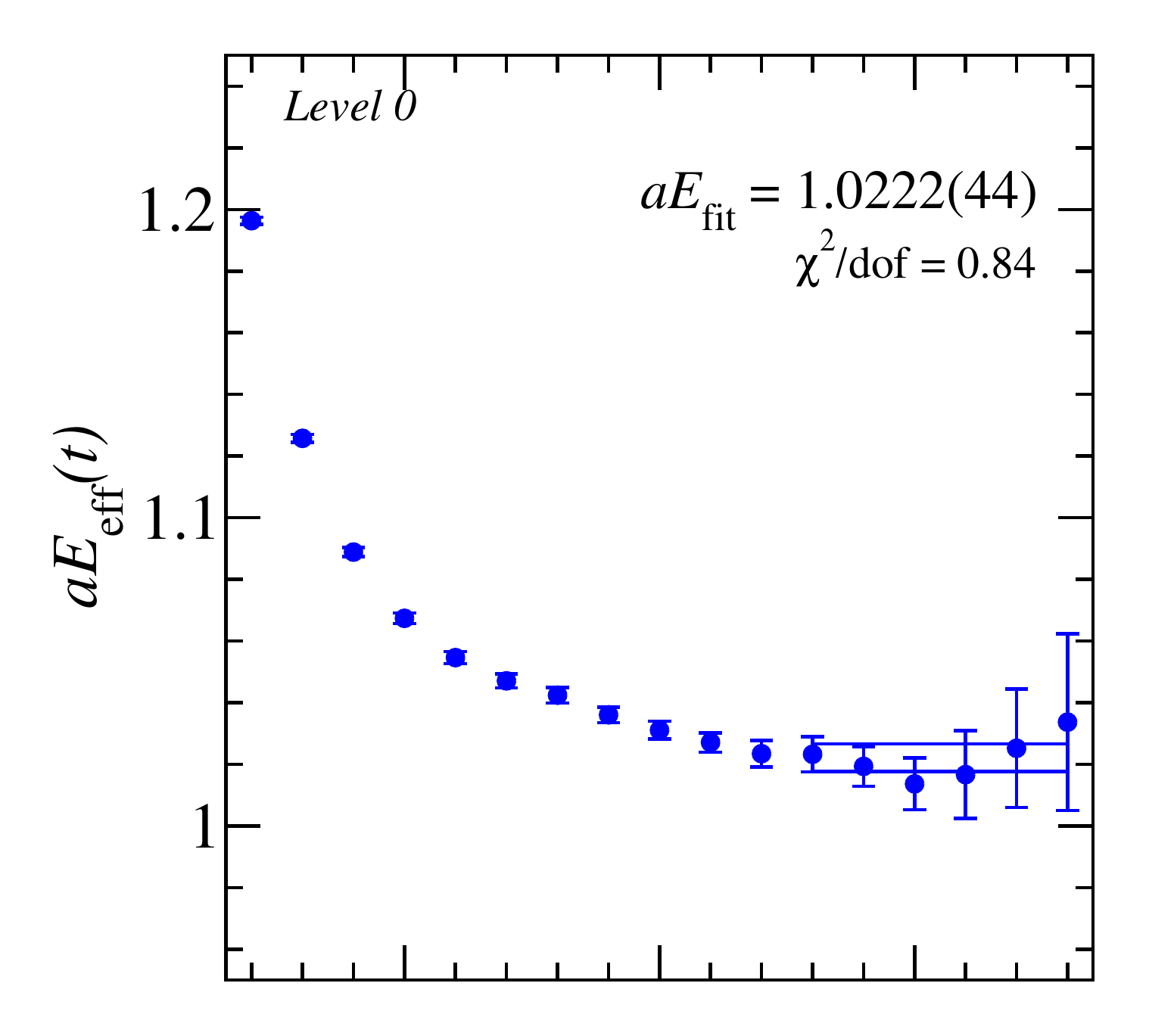}
    \hspace{-0.65cm} \includegraphics[width=0.334\textwidth]{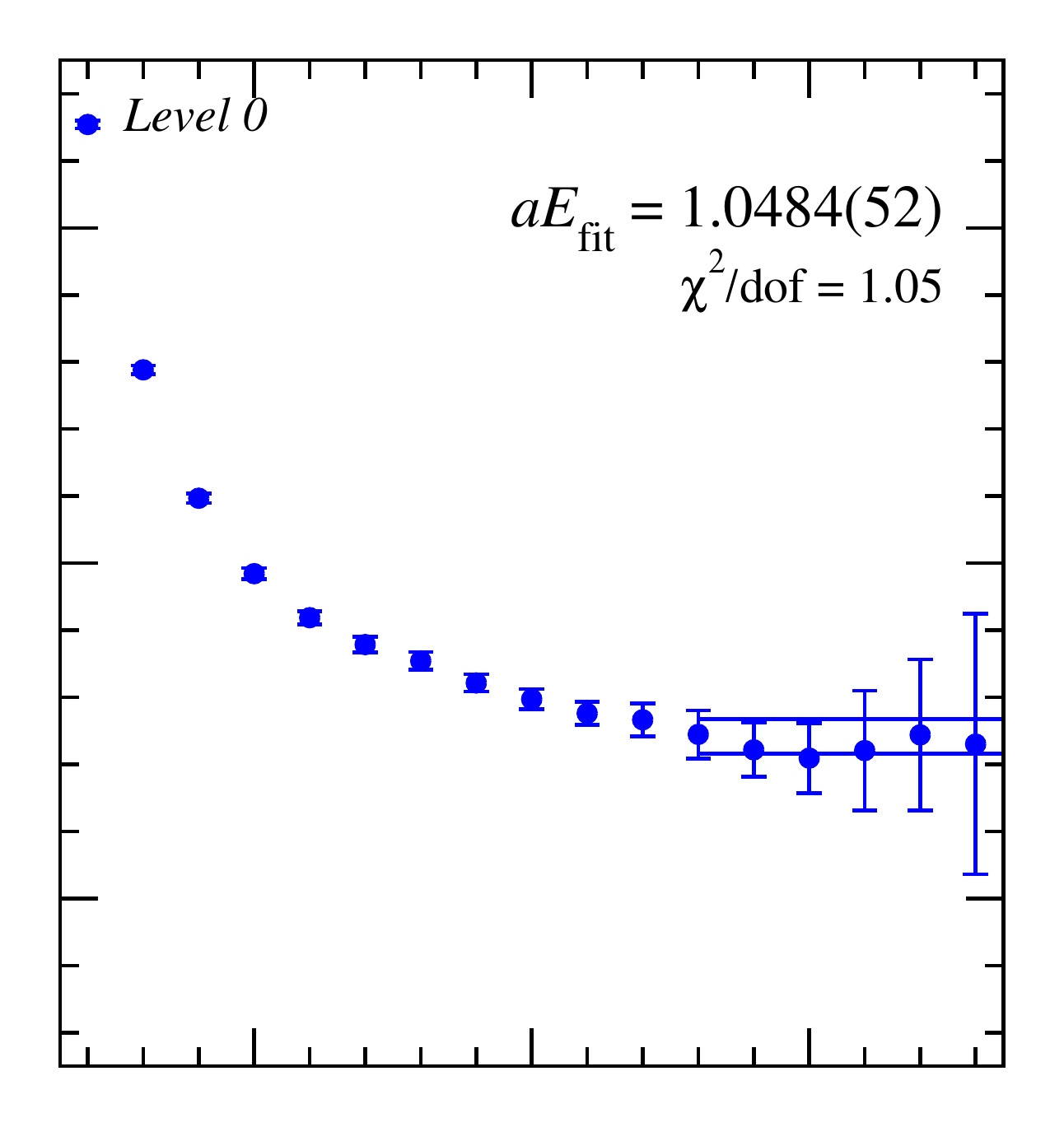}
    \hspace{-0.65cm} \includegraphics[width=0.334\textwidth]{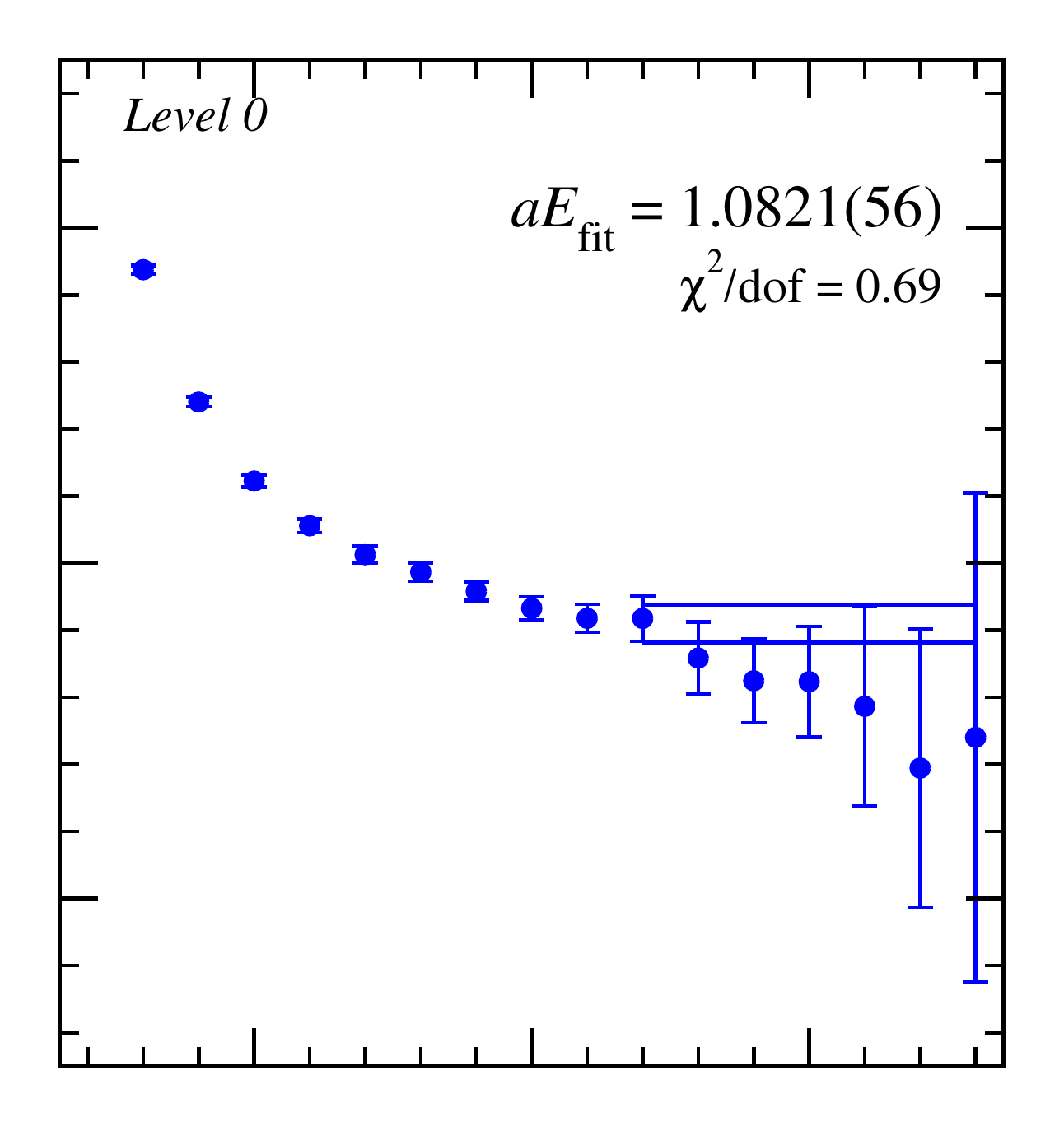}\\[-0.5cm]
                     \includegraphics[width=0.3925\textwidth]{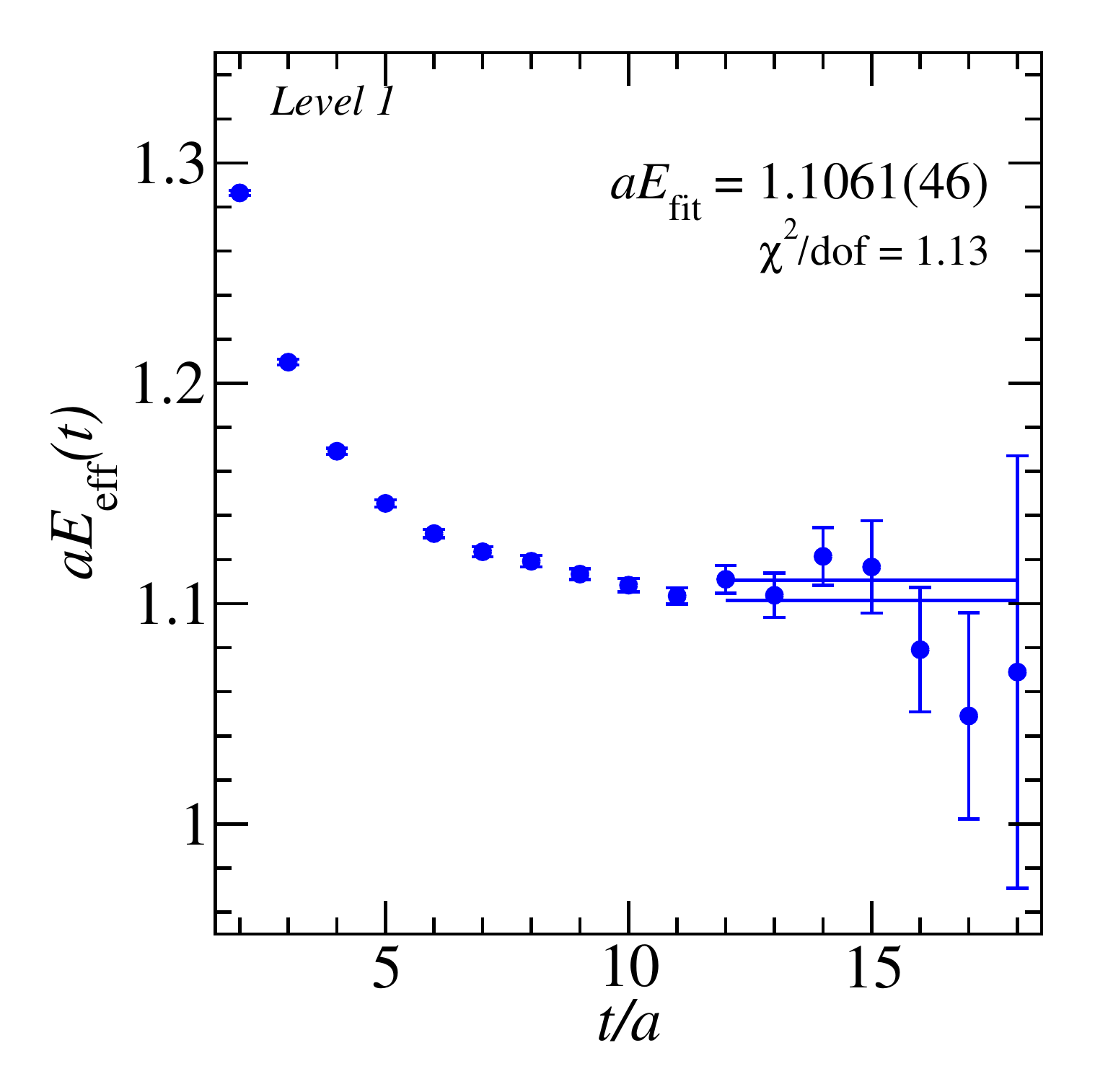}
    \hspace{-0.65cm} \includegraphics[width=0.334\textwidth]{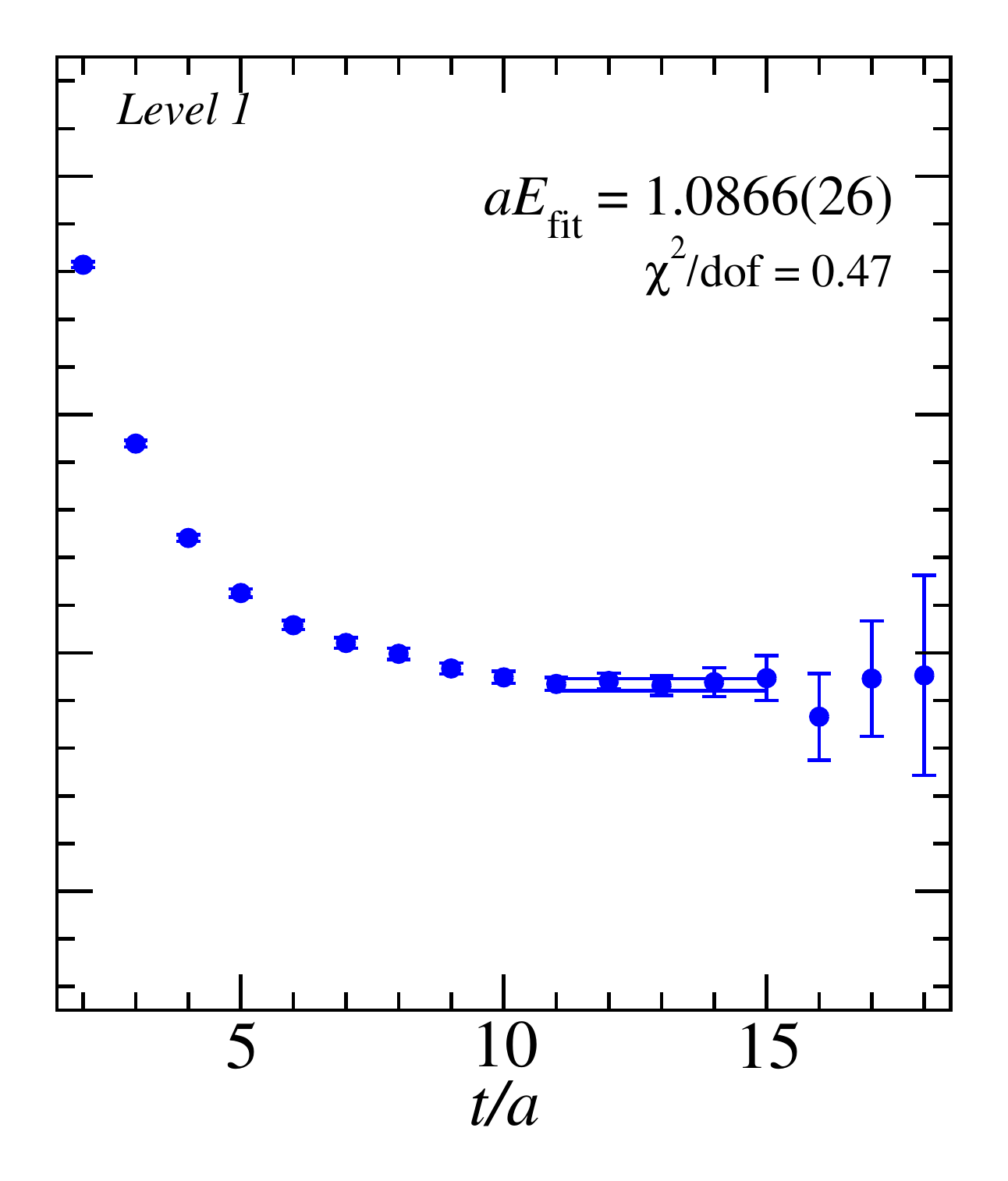}
    \hspace{-0.65cm} \includegraphics[width=0.334\textwidth]{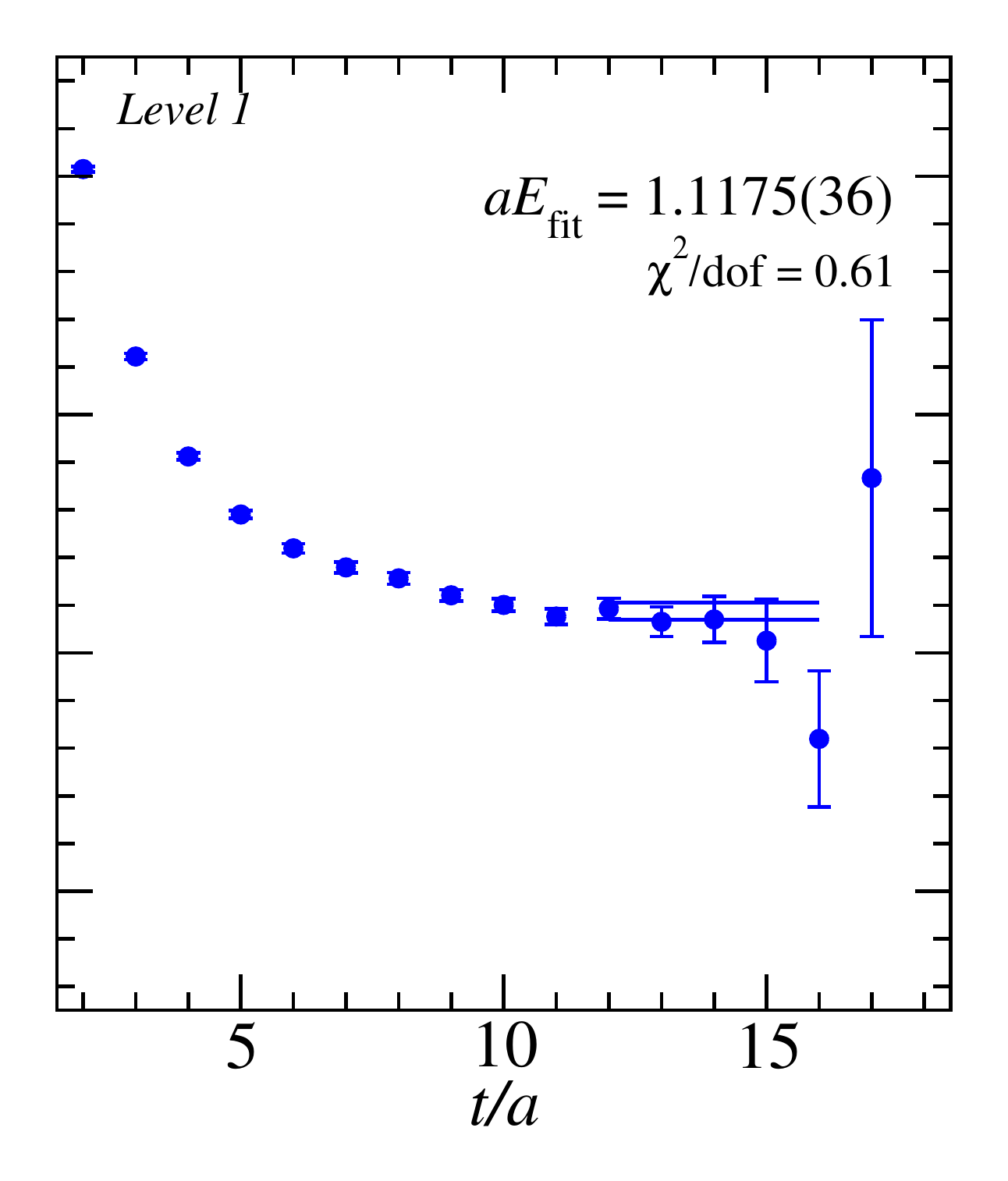}\\[-0.5cm]
    \caption{\label{fig:H101_moving_singlet}The (left) $P^2 = 1$, (middle) $P^2 = 2$, and (right) $P^2 = 3$ $A_1$ effective energies extracted from diagonalized $2 \times 2$ correlation matrices for each $P^2$ on H101.
      The basis of operators used includes spin-0 flavor-symmetric operators transforming in the $\bm{1}$-dimensional irrep of $SU(3)$.}
  \end{center}
\end{figure}

\begin{figure}[t]
  \begin{center}
                     \includegraphics[width=0.3925\textwidth]{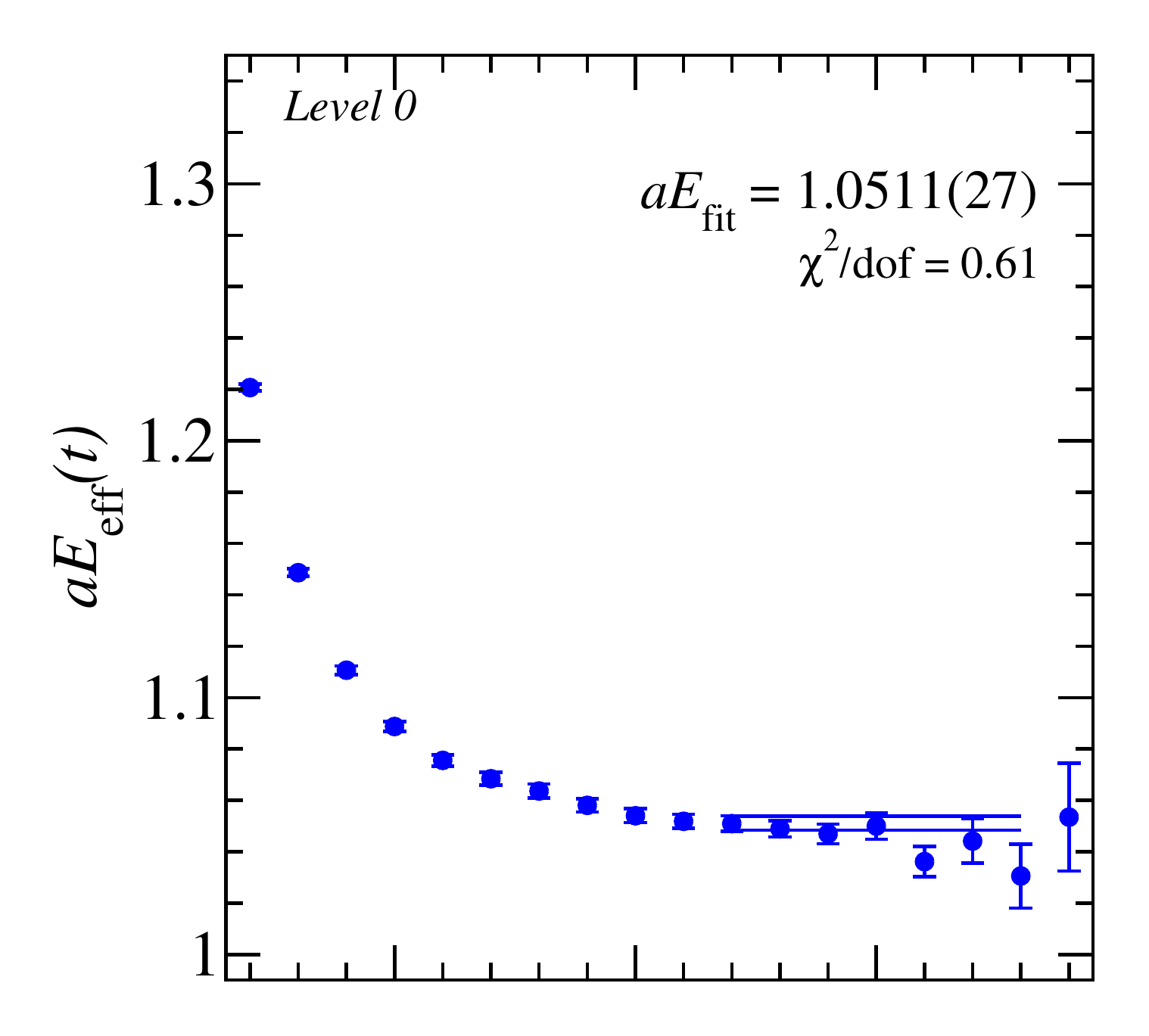}
    \hspace{-0.65cm} \includegraphics[width=0.334\textwidth]{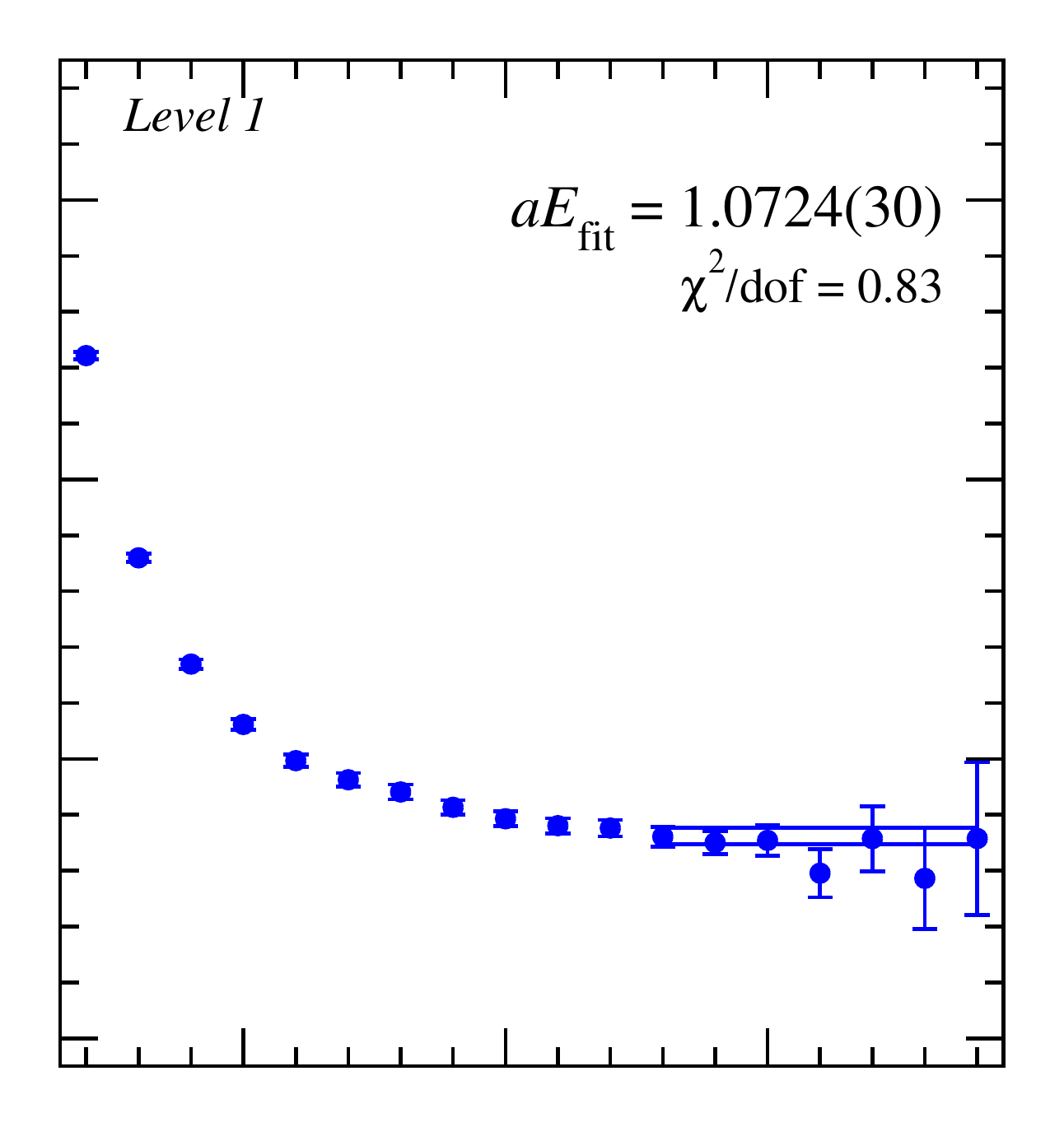}
    \hspace{-0.65cm} \includegraphics[width=0.334\textwidth]{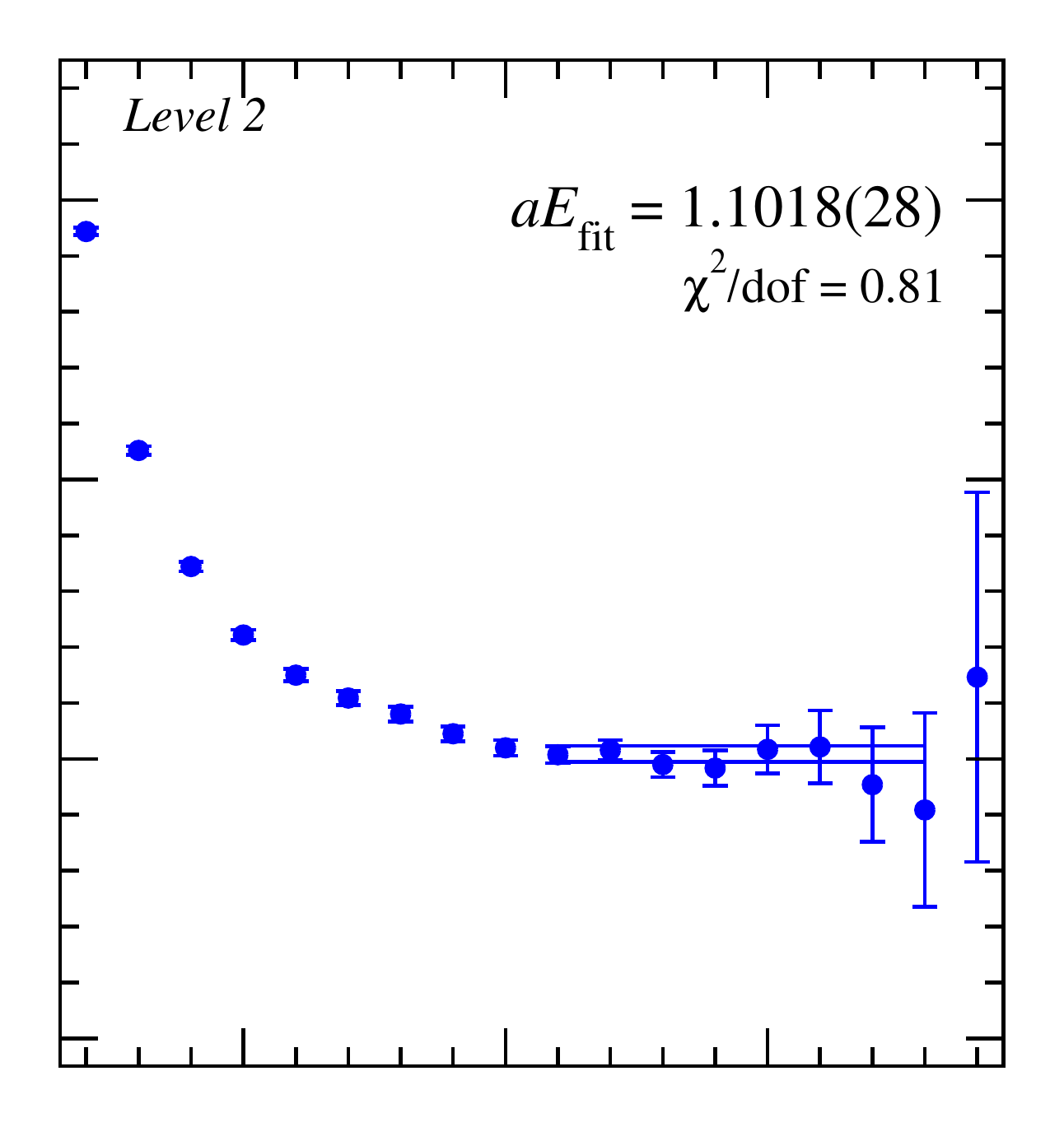}\\[-0.5cm]
                     \includegraphics[width=0.3925\textwidth]{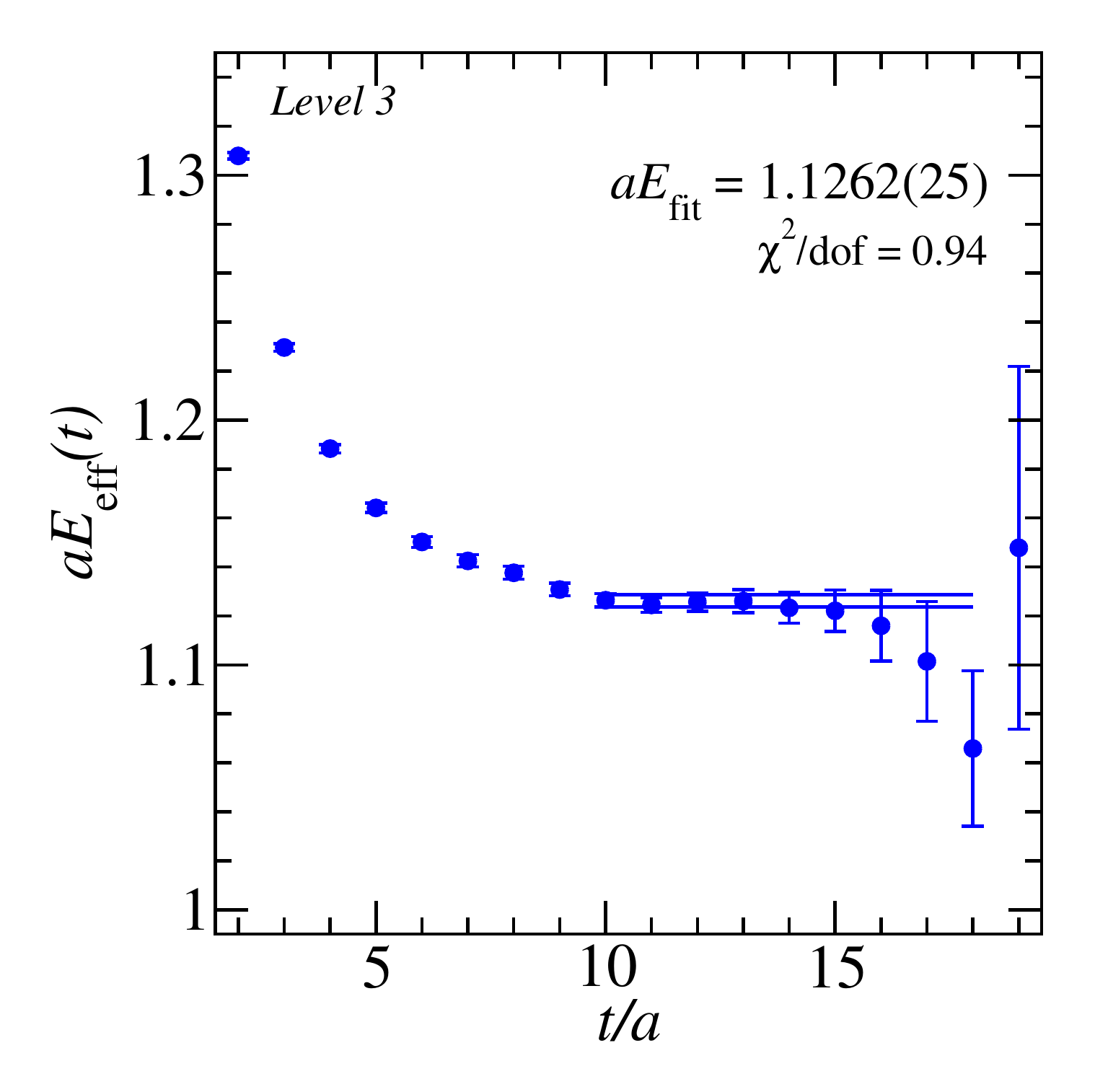}
    \hspace{-0.65cm} \includegraphics[width=0.334\textwidth]{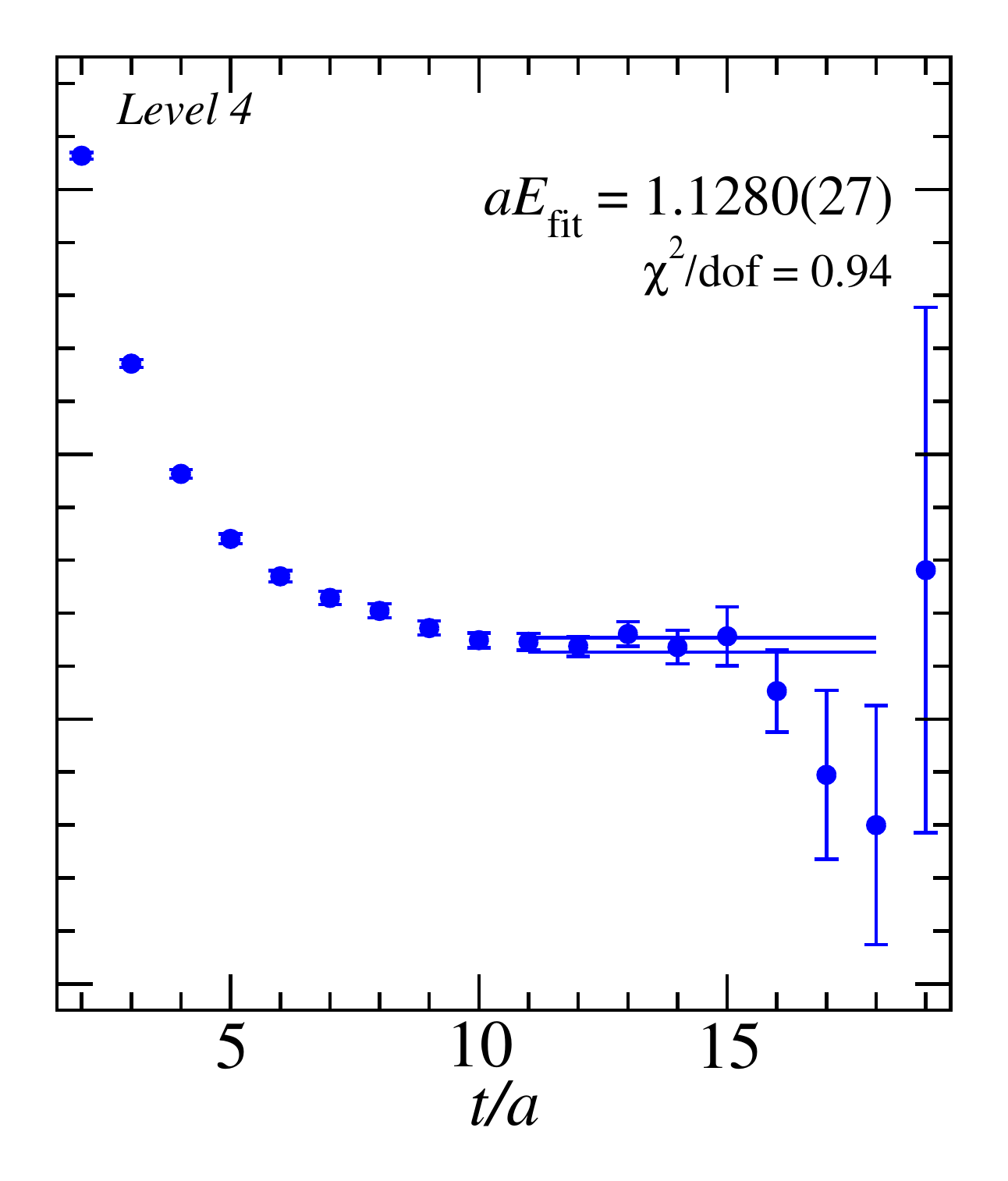}
    \hspace{-0.65cm} \includegraphics[width=0.334\textwidth]{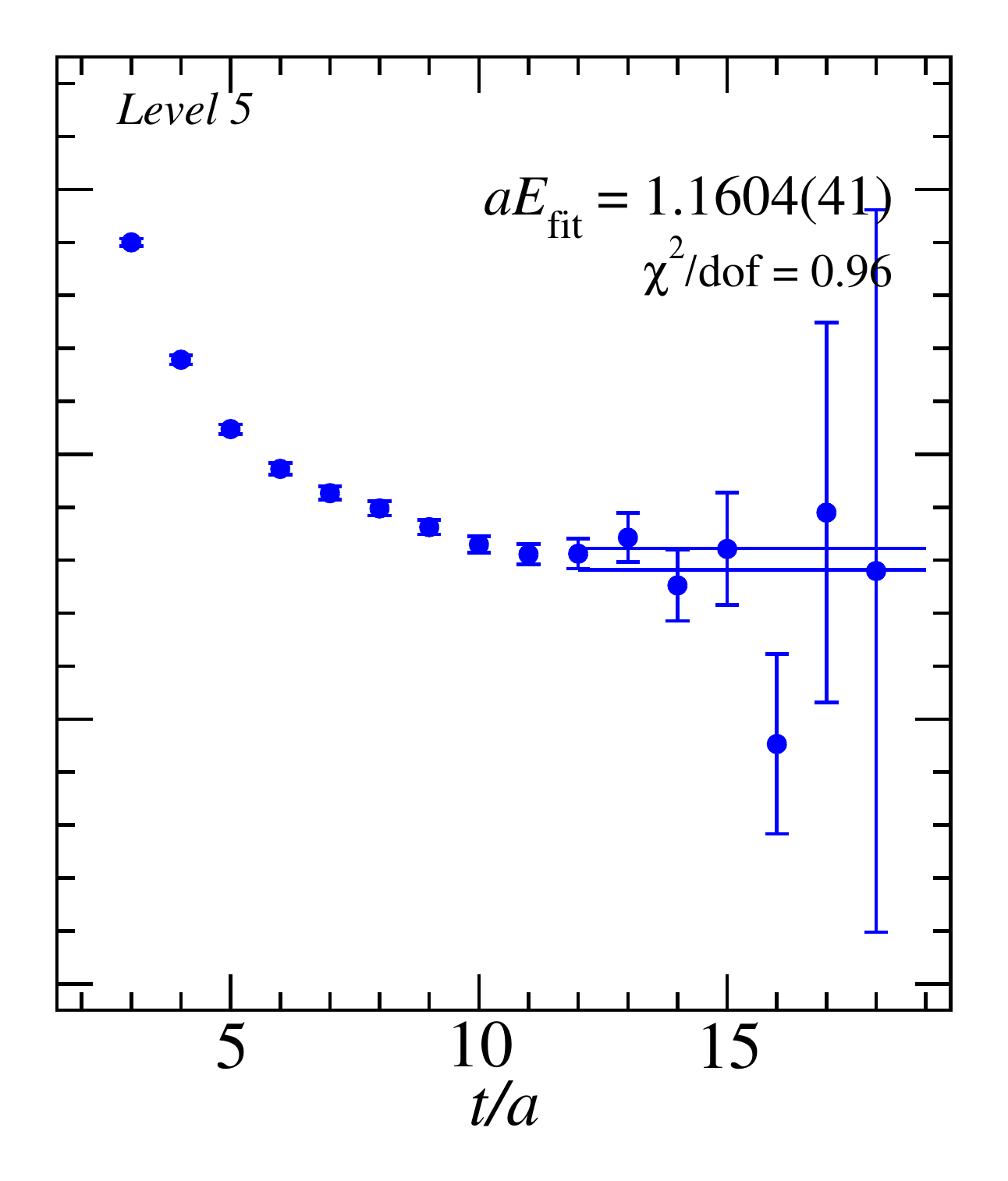}\\[-0.5cm]
    \caption{\label{fig:H101_P1_octet}The $P^2 = 1$, $A_1$ effective energies extracted from a diagonalized $6 \times 6$ correlation matrix on H101.
      The basis of operators used includes symmetric and anti-symmetric combinations of spin-0 and spin-1 operators transforming in the $\bm{8}$-dimensional irrep of $SU(3)$.}
  \end{center}
\end{figure}

The distillation method allows us to obtain estimates for two-point temporal correlation functions involving the two-baryon operators introduced in the previous section.
We briefly outline the data analysis techniques used to extract the finite-volume spectrum from these correlators.
These methods have become standard practice in the field of spectroscopy
and proceed by forming a generalized eigenvalue problem (GEVP) from a matrix of correlation functions as described in Ref.~\cite{GEVP}.
The correlation matrices are formed from each set of operators that all share the same quantum numbers\footnote{
  In the $SU(3)$-broken setup, states from different flavor multiplets can in general mix,
  but in this work, since $SU(3)$ flavor is a good symmetry, each correlation matrix will correspond to a definite flavor irrep.}
and have the form
\begin{equation}
  C_{ij} (t) = \braket{\mathcal{O}_i (t+t_0) \mathcal{O}_j^\dagger (t_0)} = \sum_{n=0}^{\infty} \bra{0} \mathcal{O}_i \ket{n} \bra{0} \mathcal{O}_j \ket{n}^\ast e^{-E_n t},
  \label{eq:correlator}
\end{equation}
where the index on the operators label all properties of the given operator.
The GEVP can be reformulated as an ordinary eigenvalue problem of the form
\begin{equation}
  C^{-1/2} (\tau_0) C(t) C^{-1/2} (\tau_0) \upsilon_n (t, \tau_0) = \lambda_n (t, \tau_0) \upsilon_n (t, \tau_0) ,
  \label{eq:gevp}
\end{equation}
where $n$ labels the different eigenvectors/eigenvalues and $\tau_0$ is referred to as the metric time.
The eigenvalues can be shown to behave as $\lambda_n (t, \tau_0) \propto e^{-E_n t}$ to leading order~\cite{GEVP}.
As an initial analysis, in order to avoid the complications involved with performing eigenvector pinning,
a diagonalization is performed at a particular time $t=\tau_D$, 
and then the eigenvectors at $\tau_D$ are used to rotate $C^{-1/2} (\tau_0) C(t) C^{-1/2} (\tau_0)$ for all other times.
Finally, single-exponential correlated fits to the diagonal elements are used to extract the energies $E_n$ in each channel.

The ground state energies extracted using spin-0 operators at rest for each ensemble
and spin-1 operators at rest for U103 are shown in Fig.~\ref{fig:ground_states}.
We have also calculated correlators in moving frames for each ensemble.
Diagonalized effective energies on H101 in three moving channels relevant for the $H$ dibaryon are shown in Fig.~\ref{fig:H101_moving_singlet}.
One can also construct both flavor-symmetric and anti-symmetric operators that transform in the $\bm{8}$-dimensional irrep of $SU(3)$,
which effectively enlarges the basis of operators in this case.
Thus we are able to extract six energies after diagonalization using these operators in the $P^2=1$ frame, which are shown in Fig.~\ref{fig:H101_P1_octet} for H101.
For all these results, values of $\tau_0/a = 8$ and $\tau_D/a = 12$ were chosen.
It was explicitly verified that the rotated correlator matrices remained diagonal for $t > \tau_D$,
but systematic errors due to choices of the operator basis, the metric time, the diagonalization time, \etc have not yet been assessed.
For instance, it was determined in Ref.~\cite{GEVP} that if $\tau_0$ is chosen sufficiently large,
then the leading order correction to the $n$th eigenvalue falls off much faster than one might naively expect,
and thus one can improve their results dramatically by properly choosing $\tau_0$.

\section{Discussion and Outlook}

The distillation method has allowed for a very precise extraction of the energy spectrum by solving a GEVP for correlation matrices in multiple channels.
The results shown in Fig.~\ref{fig:ground_states} naively indicate the presence of a bound state in the $H$-dibaryon channel,
even as the volume increases, but a full L\"uscher analysis is necessary to properly assess the finite-volume corrections.
Further, in some cases, excited-state contamination appears to be prevalent,
as can be seen, for instance, by the persisting downward trend in the top right panel of Fig.~\ref{fig:H101_moving_singlet}.
It might be possible to suppress the leading order corrections to the diagonalized correlators by better choices of the metric time $\tau_0$,
or by performing two-exponential fits.
All of these issues, including effects from other sources of systematic error, are being pursued presently.

It has also been demonstrated from chiral effective field theory that $SU(3)$ breaking effects are significant and must be properly considered~\cite{su3_breaking_1, su3_breaking_2}.
Hence, we have begun calculations on $N_f = 2 + 1$ CLS ensembles away from the $SU(3)$ symmetric point.
Of course, this involves a more complicated analysis due to the coupled $\Lambda \Lambda$-$\Sigma \Sigma$-$N \Xi$ channels.
And, as we begin to analyze the $J^P = 1^+$ sector using the spin-1 flavor anti-symmetric operators, the inclusion of multiple partial waves becomes more important due to the physical $^3S_1$-$^3D_1$ mixing.

\acknowledgments

AH would like to thank Ben H\"orz and Daniel Mohler for helpful discussions and also Colin Morningstar for providing the SigMonD analysis suite used in this work.
Our calculations have been performed at the John von Neumann Institue for Computing, J\"ulich (project HMZ21).
We are also grateful to our colleagues in the CLS initiative for sharing ensembles.

\bibliographystyle{JHEP}
\bibliography{references}

\providecommand{\href}[2]{#2}\begingroup\raggedright\begin{thebibliography}{10}

\bibitem{dibaryon_review}
H.~Clement, \href{https://doi.org/10.1016/j.ppnp.2016.12.004}{\emph{Prog. Part.
  Nucl. Phys.} {\bfseries 93} (2017) 195}.

\bibitem{jaffe_prediction}
R.~L. Jaffe, \href{https://doi.org/10.1103/PhysRevLett.38.195}{\emph{Phys. Rev.
  Lett.} {\bfseries 38} (1977) 195}.

\bibitem{experiment}
{\scshape KEK-E176, E373, and J-PARC-E07} collaboration,
  K.~Nakazawa, \href{https://doi.org/10.1016/j.nuclphysa.2010.01.195}{\emph{Nucl.
  Phys.} {\bfseries A835} (2010) 207}.

\bibitem{NPLQCD_Nf3}
{\scshape NPLQCD} collaboration, S.~R. Beane
  et~al., \href{https://doi.org/10.1103/PhysRevD.87.034506}{\emph{Phys. Rev.}
  {\bfseries D87} (2013) 034506}.

\bibitem{HALQCD_Nf3}
{\scshape HAL QCD} collaboration, T.~Inoue
  et~al., \href{https://doi.org/10.1103/PhysRevLett.106.162002}{\emph{Phys. Rev.
  Lett.} {\bfseries 106} (2011) 162002}.

\bibitem{NPLQCD_Nf21}
S.~R. Beane et~al., \href{https://doi.org/10.1142/S0217732311036978}{\emph{Mod.
  Phys. Lett.} {\bfseries A26} (2011) 2587}.

\bibitem{HALQCD_Nf21}
{\scshape HAL QCD} collaboration, K.~Sasaki
  et~al., \href{https://doi.org/10.1051/epjconf/201817505010}{\emph{EPJ Web
  Conf.} {\bfseries 175} (2018) 05010}.

\bibitem{mainz_2flavor}
A.~Francis, J.~R. Green, P.~M. Junnarkar, C.~Miao, T.~D. Rae and H.~Wittig,
  \href{https://arxiv.org/abs/1805.03966}{{\ttfamily 1805.03966}}.

\bibitem{luescher}
M.~L\"uscher, \href{https://doi.org/10.1016/0550-3213(91)90366-6}{\emph{Nucl.
  Phys.} {\bfseries B354} (1991) 531}.

\bibitem{luescher_moving}
K.~Rummukainen and S.~A.
  Gottlieb, \href{https://doi.org/10.1016/0550-3213(95)00313-H}{\emph{Nucl.
  Phys.} {\bfseries B450} (1995) 397}.

\bibitem{ensembles}
M.~Bruno et~al., \href{https://doi.org/10.1007/JHEP02(2015)043}{\emph{JHEP}
  {\bfseries 02} (2015) 043}.

\bibitem{distillation}
{\scshape Hadron Spectrum} collaboration, M.~Peardon
  et~al., \href{https://doi.org/10.1103/PhysRevD.80.054506}{\emph{Phys. Rev.}
  {\bfseries D80} (2009) 054506}.

\bibitem{su3}
{\scshape HAL QCD} collaboration, T.~Inoue
  et~al., \href{https://doi.org/10.1143/PTP.124.591}{\emph{Prog. Theor. Phys.}
  {\bfseries 124} (2010) 591}.

\bibitem{GEVP}
B.~Blossier, M.~Della~Morte, G.~von Hippel, T.~Mendes and
  R.~Sommer, \href{https://doi.org/10.1088/1126-6708/2009/04/094}{\emph{JHEP}
  {\bfseries 04} (2009) 094}.

\bibitem{su3_breaking_1}
J.~Haidenbauer and U.-G.
  Mei{\ss}ner, \href{https://doi.org/10.1016/j.physletb.2011.10.070}{\emph{Phys.
  Lett.} {\bfseries B706} (2011) 100}.

\bibitem{su3_breaking_2}
J.~Haidenbauer and U.-G.
  Mei{\ss}ner, \href{https://doi.org/10.1016/j.nuclphysa.2012.01.021}{\emph{Nucl.
  Phys.} {\bfseries A881} (2012) 44}.

\end{thebibliography}\endgroup

\end{document}